# Comb-resolved spectroscopy with immersion grating in long-wave infrared

Kana Iwakuni,[1,*] Thinh Q. Bui,[1] Justin Niedermeyer,[1] Takashi Sukegawa,[2] and Jun Ye[1]

[1]JILA, National Institute of Standards and Technology and University of Colorado, and Department of Physics, University of Colorado, Boulder, CO 80309, USA
[2]Canon Inc., 20-2, Kiyohara-Kogyodanchi,Utsunomiya, Tochigi, 321-3292, Japan
*Corresponding author: kana.iwakuni@jila.colorado.edu

**We have developed a dispersive spectrometer using a compact immersion grating for direct frequency comb spectroscopy in the long-wave infrared region of 8-10 μm. A frequency resolution of 463 MHz is achieved, which is the highest reported in this wavelength region with a dispersive direct frequency comb spectrometer. We also demonstrate individual mode-resolved imaging of the frequency comb spectrum by Vernier cavity filtering and apply this to obtain both simple and complex molecular spectra. These results indicate that the immersion grating spectrometer offers the next advancement for sensitive, high-resolution spectroscopy of transient and large/complex molecules when combined with cavity enhancement and cooling techniques.**

## 1. INTRODUCTION

### A. Mid-IR Direct Frequency Comb Spectroscopy

Direct frequency comb spectroscopy (DFCS) in the mid-infrared spectral region find important applications in both fundamental laboratory spectroscopy and remote sensing. In particular, mid-IR DFCS is an attractive approach for molecular spectroscopy because of the afforded high sensitivity, broad spectral coverage, and rapid acquisition capabilities. The fast acquisition advantage allows for both real-time remote sensing [1] (seconds) and measurement of fast chemical kinetics [2-5] (microseconds). These combined advantages have been demonstrated in techniques of cavity-enhanced spectroscopy with a dispersive spectrometer [2-4, 6]. As alternatives to a dispersive spectrometer, mid-IR DFCS has also been demonstrated with dual-comb spectroscopy [5, 7-10] and Fourier transform spectroscopy [11-13].

In a dispersive spectrometer, the frequency comb spectrum is spatially dispersed in two spatial dimensions (2D) by a combination of a virtually imaged phase array (VIPA) etalon and a reflective grating, and then imaged onto a camera [14]. In this case, the VIPA spectrometer can record spectra within 10 μs, limited by the integration time of the camera, which is required to observe transient chemical species [2-5]. Unfortunately, the VIPA etalon has the inherent disadvantage of a limited-bandwidth due to optical coatings, as well as a relatively low throughput of ~20 %. The former is problematic for very broadband spectroscopy, and each etalon design must be tailor-made to the specific wavelength application.

With VIPA spectrometers, DFCS has been demonstrated in the mid-infrared region up to 5 μm. Relative to shorter mid-IR wavelengths (< 5 μm), important molecular targets in atmospheric sciences like the Criegee intermediate [15], NO$_3$ radical [16, 17], isoprene [18], and fundamental spectroscopy like buckyball (C$_{60}$) [19], display significantly larger absorption intensities near 10 μm. A comparably important consideration for these larger, complex molecules is that spectroscopic probing at longer infrared wavelengths alleviates spectral congestion due to IVR (intramolecular vibrational redistribution) processes, thus enabling quantum-state resolution [6, 20]. These advantages motivate the recent construction of an 8-10 μm mid-infrared frequency comb (an optical parametric oscillator (OPO) based on AgGaSe$_2$) for DFCS [21]. Here, we report a dispersive spectrometer for 8-10 μm DFCS comprised of an immersion grating, a reflective grating, and a strained layer superlattice (SLS) detector element camera for imaging. The enabling technology for this spectrometer is the immersion grating, which has been fabricated for wavelengths > 5 μm.

### B. Immersion Grating

An immersion grating is essentially an echelle grating with a large blaze angle. However, unlike traditional echelle gratings, the diffraction surface is immersed in a high refractive index ($n$) material, so that the angular dispersion of the immersion grating is enhanced by $n$. Therefore, the physical size of immersion grating can be reduced by a factor of $n$ to achieve the same resolving power as a reflective echelle grating. According to the Rayleigh criterion, the resolving power ($RP$) of an immersion grating is described as

$$RP = \lambda/\Delta\lambda = (2nw\tan\theta)/\lambda , \qquad (1)$$

where $w$ is the diameter of the input beam, $\theta$ is the blaze angle, and $\lambda$ is the wavelength. $w \times \tan\theta$ indicates the effective beam diameter illuminating grooves and the resolving power is proportional to the input beam diameter. Although the immersion grating concept has been around for some time [22, 23] for silicon [24-26], ZnSe [27], GaP [27], and bismuth germinate [27], recent advances in machining techniques of brittle crystals like CdZnTe [28], InP [29], and germanium have enabled their realization [29-31] and potential application for spectroscopy in the infrared. In the field of infrared astronomy, immersion gratings are highly desired for their high resolving power, which for a normal echelle grating would have been of much larger sizes. Immersion gratings have been already installed on international telescopes [32-34].

The immersion grating used in this work is fabricated from a single germanium crystal. The Ge immersion grating can achieve relatively high diffraction efficiency comparable to a traditional reflective grating, and its resolving power can be arbitrarily scaled by increasing the input beam size. Therefore, this immersion grating is a versatile candidate for use in a high resolution, long mid-infrared spectrometer due to its transparency from 2 μm to 16 μm and large reflective index ($n$ = 4). A preliminary demonstration of DFCS with a Ge immersion grating was reported at 4 μm [35]. Here, we apply this grating for DFCS in the 8-10 μm spectral region. In this study, we report in-depth characterization of this dispersive spectrometer, with focused discussions on diffraction efficiency, resolving power, noise performance, its use for high resolution molecular spectroscopy, and the potential for time-resolved spectroscopy.

## 2. APPARATUS AND CHARACTERIZATION

### A. Immersion grating spectrometer

Figure 1(a) shows an overview of the immersion grating spectrometer. Our Ge immersion grating has dimensions of 49 × 41 × 155 mm as shown in the top panel in Fig.1 (a). The blaze angle is 75 ° and the lattice constant (groove spacing) is 476 μm. The diffraction surface is Au-coated to maximize the diffraction efficiency. The input facet of the immersion grating is AR-coated to cover the whole spectrum of the light source, which is OPO with wavelength tunable 8-10 μm [21] (about 200 nm). The incident light enters the immersion grating at an angle about normal to the surface, which corresponds to the blaze angle (75 degrees). Due to the large blaze angle and groove spacing characteristic of echelle-type gratings as well as high refractive index of Ge, the diffraction order, $m$, is high (~ 432) and thus are highly spatially overlapped. The resulting grating free spectral range (FSR), given by $\lambda/m$ ~ 20 nm, is much narrower than the spectral bandwidth of the OPO. A conventional reflective grating is inserted as an orthogonal cross-disperser to map the full frequency comb spectrum onto a 2D image. Since the cross-disperser grating only needs to separate wavelengths covering one FSR (20 nm) set by the immersion grating, its required resolving power is relatively low, at about 500. Our cross disperser grating has a groove spacing of 13 μm and an overall dimension of 68 × 68 × 9 mm. Finally, the diffraction efficiency is 50 % for the immersion grating and 70 % for the cross disperser grating in a quasi-Littrow configuration.

Figure 1(b) shows a typical 2D image taken by the $LN_2$-cooled SLS camera with 640 × 512 pixels, 20-μm pixel pitch, and frame rate of 100 Hz. The frequency comb wavelength is vertically dispersed by the immersion grating and horizontally dispersed by the cross disperser grating into a series of vertical fringes. The image for each fringe is not comb-tooth resolved. According to Eq. (1), the calculated frequency resolution for a beam diameter ($1/e^2$ diameter) of 18 mm is about 570 MHz, which is larger than typical repetition rate of our fiber-based frequency comb ($f_{rep}$ = 110 MHz), such that mode filtering is required to observe a comb mode-resolved image. The width of one fringe is about 90 μm (4.5 pixels) which is consistent with the expected beam diameter from Gaussian beam propagation through the imaging optics after the gratings. The separation between fringes, which is determined by the angular dispersion of the cross disperser grating and the focal length of the imaging lenses, is about 280 μm (14 pixels), which is more than sufficient to avoid inter-fringe crosstalk. The entire OPO spectrum is mapped on 28 fringes that cover approximately two-thirds of the camera detector size.

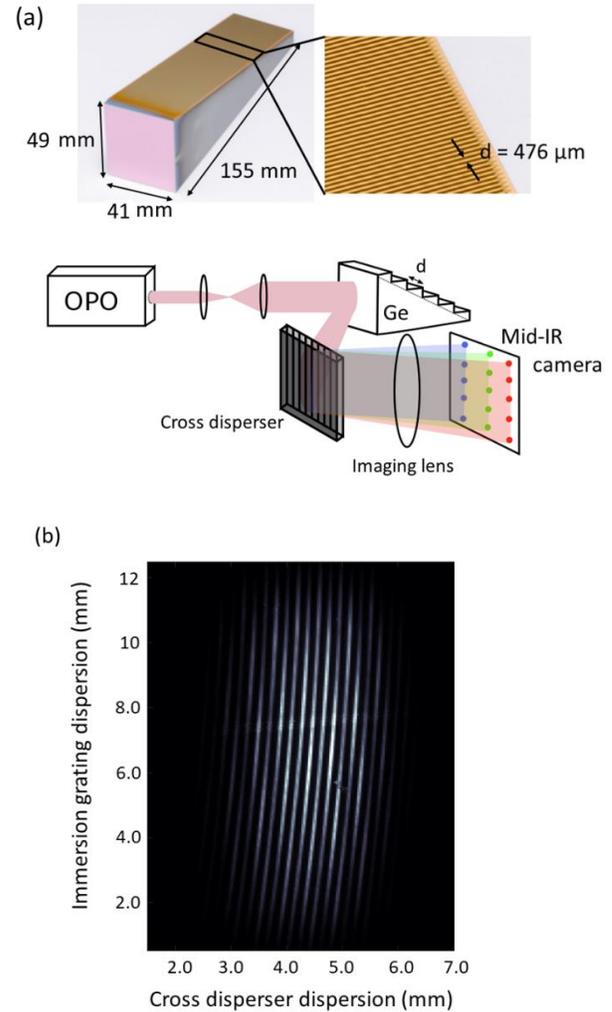

**Fig. 1.** DFCS spectrometer. (a) A schematic of the Ge immersion grating and an overview of the immersion grating spectrometer. $d$ is the lattice constant of the immersion grating. The OPO is an optical parametric oscillator operating at 8-10 μm. The diffracted light from the Ge immersion grating is cross-dispersed with a reflective grating and mapped onto a camera as a 2D image. (b) Camera image of the dispersed comb light. The comb modes are dispersed vertically by the immersion grating and horizontally by the cross disperser grating.

### B. Grating Free Spectral Range (FSR)

We characterize the FSR of the immersion grating from measurement of molecular absorption features. An absorption cell is inserted before the beam expansion lenses shown in Fig. 1(a). Figure 2(a) shows recorded images of rotationally-resolved $D_2O$ and $N_2O$ absorption of the immersion grating spectrometer. In Fig. 2(a), a negative image is shown to accentuate molecular absorption indicated by bright dots. The FSR is determined by locating the repeating absorption patterns, which is most apparent in the $D_2O$ image. The red boundary lines in Fig. 2(a) indicate one FSR. The $N_2O$ spectrum also corroborates this observation. Since the wavelength spacing between rotational transitions of $N_2O$ is fairly constant at 6 nm (~ 0.8 $cm^{-1}$), and one FSR is calculated to by ~ 20 nm (section 2A), we would expect approximately 4 absorption dots to span one FSR. This is consistent with the data in the absorption image of $N_2O$ by counting the number of absorption dots on a single fringe. To obtain a traditional frequency-domain spectrum, the fringes are rastered from top to bottom and left to right by a fringe-finding algorithm. To avoid redundancy in the overlapping diffraction orders, the image is first cropped to discover fringes only within one FSR region.

## C. Molecular Spectroscopy

Once the grating FSR has been determined, molecular absorption can be quantitatively determined from $(I_0 - I)/I_o$, where $I$ and $I_0$ are the camera images with (signal image) and without (reference image) absorption, respectively. Figure 2(b) shows the observed single-shot, absorption spectra of $N_2O$, $D_2O$, and dimethyl ether retrieved from the fringe-finding algorithm. The frequency axes are calibrated with $N_2O$ and $D_2O$ rovibrational spectra reported in the HITRAN database [36].

In contrast to the 3-5 μm where predominately vibrational stretching fundamentals are observed, the 8-10 μm region provides access to lower frequency bending modes, such as those observed in the methyl rocking motion of dimethyl ether centered at ~ 1175 cm$^{-1}$ (8510 nm). Dimethyl ether is an important molecule in astrophysics and the interstellar medium, and is also one of the simplest molecules displaying large-amplitude motions, corresponding to the internal rotations of the two methyl groups [37]. A combination of the tunneling splitting of the rotational levels caused by the methyl torsional motion and highly perturbed nature of the vibrational excited state results in a complex and intractable spectrum, which could be resolved if the molecules were rotationally and vibrationally cooled, for example by buffer gas cooling or molecular beam expansion. Here several room temperature dimethyl ether spectra are recorded at different sample pressures. Although rovibrational assignments are not possible at the current room temperature conditions, this is the first report of a rotationally-resolved dimethyl ether spectrum of this band, which demonstrates that the immersion grating spectrometer has potential for extending the applications of high-resolution spectroscopy towards larger molecules with complex structure.

## D. Resolution

The frequency resolution of the spectrometer is set by the resolving power of the immersion grating and the imaging system. We experimentally determined the spectrometer resolution using two different approaches: 1) pressure dependent linewidth measurements of $N_2O$ absorption lines and 2) comb mode resolved imaging with a comb-mode-filtering cavity. In Fig. 3(a), the linewidth of pure $N_2O$ samples is plotted as a function of pressure. At high pressures, the linewidth of the absorption is determined by pressure broadening and its line profile is Lorentzian. At low pressures, the linewidth is determined by the instrument linewidth and Doppler broadening, which is described by a Gaussian line profile. Here, we measured the pressure-dependent linewidth of $N_2O$ at two different beam diameters incident on the immersion grating. The slope of the figure shows the pressure broadening coefficient and both slopes are about 10 MHz/Torr for $N_2O$ self-broadening, which agrees with the HITRAN database. By extrapolating back to zero pressure (y-intercept), the instrument-limited frequency resolution (FWHM) is measured to be 623(5) and 463(16) MHz with a beam diameter of 18(1) mm and 34(1) mm, respectively. A resolution of ~463 MHz is the highest reported for DFCS with a dispersive spectrometer in this wavelength region.

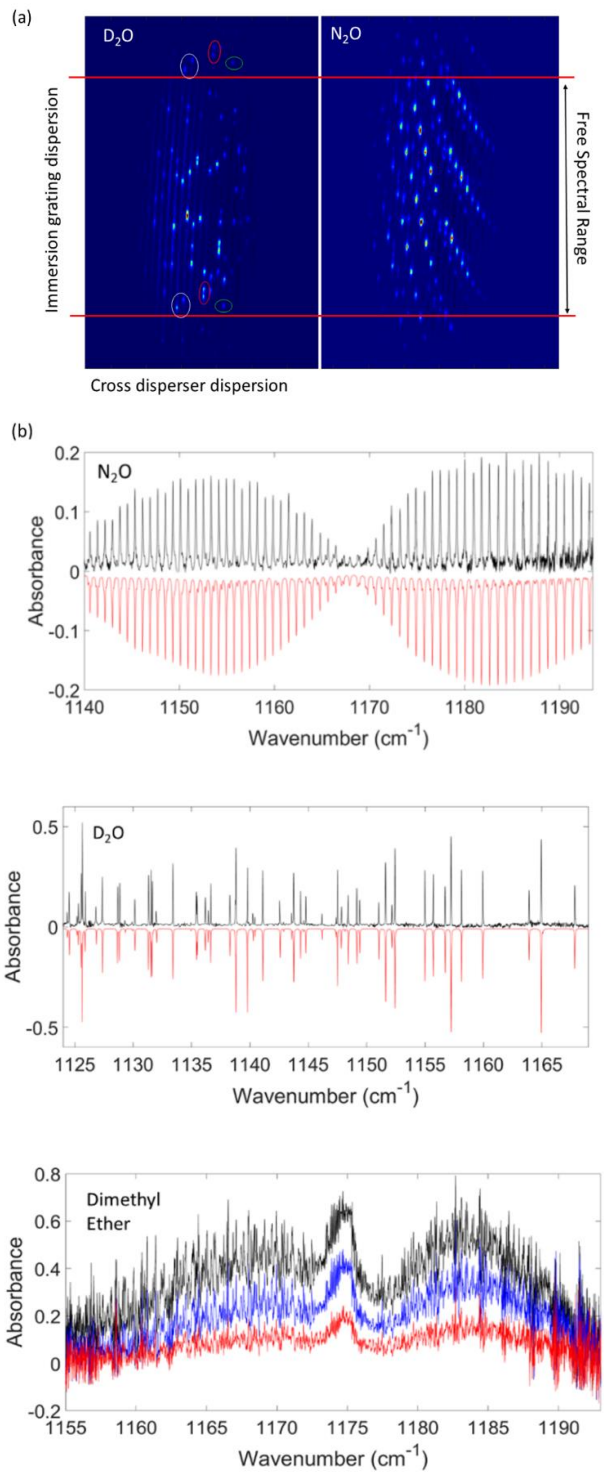

**Fig. 2.** Molecular spectroscopy. (a) Observed camera images (negative) of $D_2O$ and $N_2O$ absorption shown in bright dots. The marked absorption dots in the $D_2O$ image show the repeating pattern, indicating they are the same spectral feature. The vertical gap separated by the two horizontal red lines indicates FSR of the immersion grating. (b) Observed spectrum of $N_2O$, $D_2O$, and dimethyl ether. The frequency axis is calibrated with $N_2O$ and $D_2O$ spectra reported in the HITRAN database. The simulation is shown as inverted red traces. Absorption spectra of dimethyl ether were taken with three different sample pressures.

The expected spectrometer resolution is a convolution of the immersion grating and the imaging system instrument functions. The former was calculated using Eq. 1 in which the resolving power of the immersion grating is determined by the number of illuminated grooves, and therefore, input beam size $w$. The latter was calculated from the image mapping of the linear dispersion of the gratings and beam diameter on the camera detector plane. For input beam diameters of 18(1) and 34(1) mm, we obtained expected spectrometer resolutions of 650(100) MHz and 460(34) MHz, respectively, which are in good agreement with the measured values. In this current system, the beam size, thus resolving power, is limited by the input facet dimensions of the immersion grating, which is only limited by the fabrication process for producing larger Ge crystals.

As additional validation of the spectrometer resolution, we measured the comb mode-resolved camera image. To obtain the comb mode-resolved image, we use the cavity-filtering (Vernier) technique in which one comb mode is filtered out every 19 modes using a Fabry-Perot cavity, so that the resulting effective repetition rate is 2.09 GHz. This value is much larger than the resolution of the immersion grating, resulting in resolvable comb teeth shown in Fig. 3(b). To lock the comb laser $f_{rep}$ to the Vernier filter cavity FSR, the swept-locked technique is used [38]. Here, we sweep $f_{rep}$ by modulating the comb oscillator cavity length, and the feedback error signal is used to control the length of the ring PZT attached to one of the cavity mirrors in the Vernier cavity. The comb teeth spacing in Fig. 3(b) is the expected 2.09 GHz, and the spectrometer resolution is determined by fitting the FWHM (740 (80) MHz) of each comb tooth to a Gaussian line profile.

### E. Noise Characterization and Sensitivity

Finally, we characterized the spectrometer's sensitivity for direct absorption experiments. The intensity of transmitted light is given by Lambert-Beer Law:

$$I = I_0 \exp(-\alpha L) \approx I_0 (1 - \alpha L) \text{ for } \alpha L \ll 1. \quad (2)$$

Here, $\alpha$ is the absorption coefficient, and $L$ is the absorption length. Solving for $\alpha L$ in Eq. (2) yields

$$\alpha L \approx (I_0 - I)/I_0 \equiv \Delta I / I_0. \quad (3)$$

We determined noise processes on absorption from measurement of $\Delta I/I_0$. To characterize the noise of this measurement, 100 sets of three images were recorded every 200 ms with comb light incident on the camera sensor in the absence of absorption. The three images within each set were obtained at 250 μs integration time, each separated by 10 ms, corresponding to a frame rate of 100 Hz. The first image is a background image (B) where the laser light is blocked by a chopper and the two others ($S_0$ and $S_1$) measure the intensity of the incident light such that $S_0 - B$ corresponds to $I_0$ and $S_1 - B$ corresponds to $I$. Therefore, $\Delta I/I_0 = (S_0 - S_1)/(S_0 - B)$. For a given pixel, the standard deviation of $\Delta I/I_0$ over the 100 measurements gives an estimate for the fractional intensity noise, $\delta_I$. First, we determined $\delta_I$ as a function of laser input power. We estimate the power incident on a single pixel by first determining the gain, $g$, which has units of electron/count:

$$g = \frac{\eta}{C_{tot}} \cdot \frac{P_{tot} t_{int}}{E_{photon}}, \quad (4)$$

where $\eta$ is the quantum efficiency of the camera, $C_{tot}$ is the integrated counts on illuminated pixels, $P_{tot}$ is the total incident power, $E_{photon}$ is photon energy, and $t_{int}$ is the camera integration time. From Eq. (4), the direct conversion of the measured count to optical power for a single pixel is given by the conversion factor $a$, which has units of power/count:

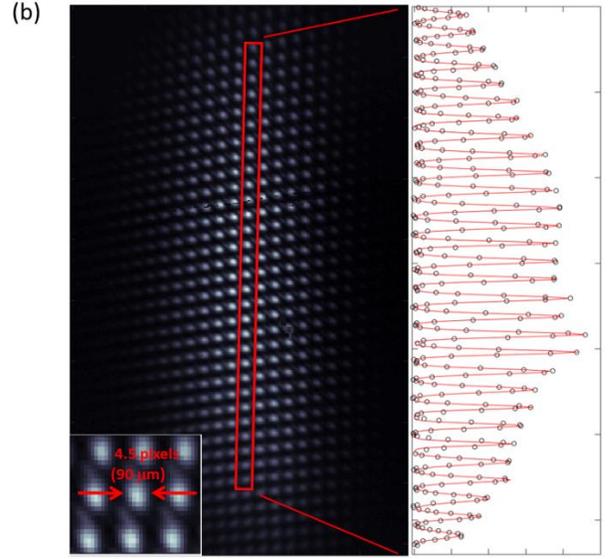

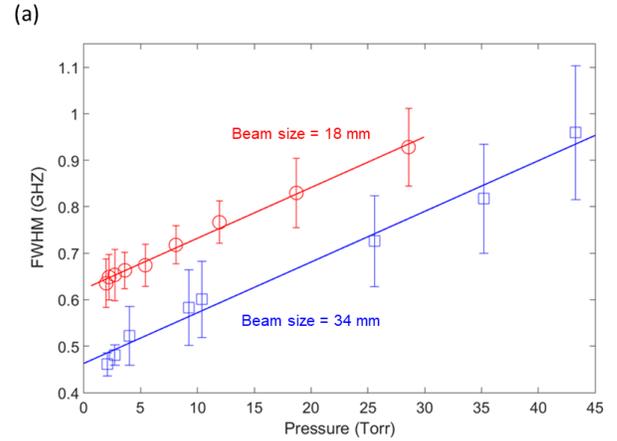

**Fig. 3.** Spectrometer resolution. (a) Pressure-dependent N$_2$O linewidth measurements with the input beam diameters of 18 mm (red) and 34 mm (blue). (b) Comb mode-resolved 2D image. One out of nineteen comb modes are filtered with a Fabry-Perot cavity (effective repetition rate is 2.02 GHz). Each comb mode is focused onto 4.5 pixels (90 μm) and displays a linewidth of about 742 MHz when the input beam diameter is 18 mm.

$$a = \frac{E_{photon}}{t_{int}} \cdot \frac{g}{\eta}. \quad (5)$$

The average $\delta_I$ for as a function of power per pixel is shown in Fig. 4(a). If no noise sources were present in the spectrometer, the minimum $\Delta I/I_0$ on a single pixel would be limited by shot noise and given by Ref. [39]

$$\left(\Delta I / I_0\right)_{min}^{shot} = \sqrt{2eB/Rp} \quad (6)$$

where $e$ is the electron charge, $B$ is the measurement bandwidth, $R$ is the responsivity of the detector and $p$ is the power incident on the pixel obtained from measure counts of single pixel multiplied by $a$ in Eq. (5). Figure 4(a) shows that $(\Delta I/I_0)_{min}^{shot}$ is roughly two orders of magnitude below $\delta_I$, indicating that we are well above the shot noise limit at the current experimental optical power.

To determine the sources of noise, we analyzed the behavior of $\delta_I$. By assuming that the major noise sources, (laser intensity noise, shot noise, and camera noise) are uncorrelated, we obtain

$$\delta_I \propto \sqrt{(A\Delta I)^2 + B^2 \Delta I + C^2}\bigg/I_0. \quad (7)$$

Laser intensity noise is proportional to $A\Delta I$, shot noise to $B(\Delta I)^{1/2}$, and camera noise (a total measurement of dark noise, readout noise, etc.) to $C$, for constants $A, B,$ and $C$. If $A\Delta I \gg B(\Delta I)^{1/2}, C$, laser intensity noise dominates; if $B(\Delta I)^{1/2} \gg A\Delta I, C$, shot noise dominates; if $C \gg A\Delta I, B(\Delta I)^{1/2}$, camera noise dominates. Until ~5 nW of incident power, the slope of the log-log plot is approximately -1, indicating that detector noise dominates. At higher powers, the $\delta_I$ becomes independent of power, indicating that laser intensity noise dominates. This observation was independently verified by an additional laser relative intensity noise (RIN) measurement.

Noise can be further averaged down. A set of 3000 measurements of $\Delta I/I_0$ were recorded with ~3 nW of total incident laser power on a pixel. $N$ measurements of $\Delta I/I_0$ were averaged together, where $N = 1,2,3,...,1500$. For a single pixel, $\delta_I$ was calculated as a function of $N$. The result for an average of 50 randomly chosen pixels is shown in Fig. 4(b). Since the slope of this line is proportional to $N^{-1/2}$, we conclude that the noise at this power is essentially random, and that we can achieve an absorption sensitivity of $\delta_I \approx 3 \times 10^{-4}$ per detection channel for 1000 measurements, or 200 s of acquisition.

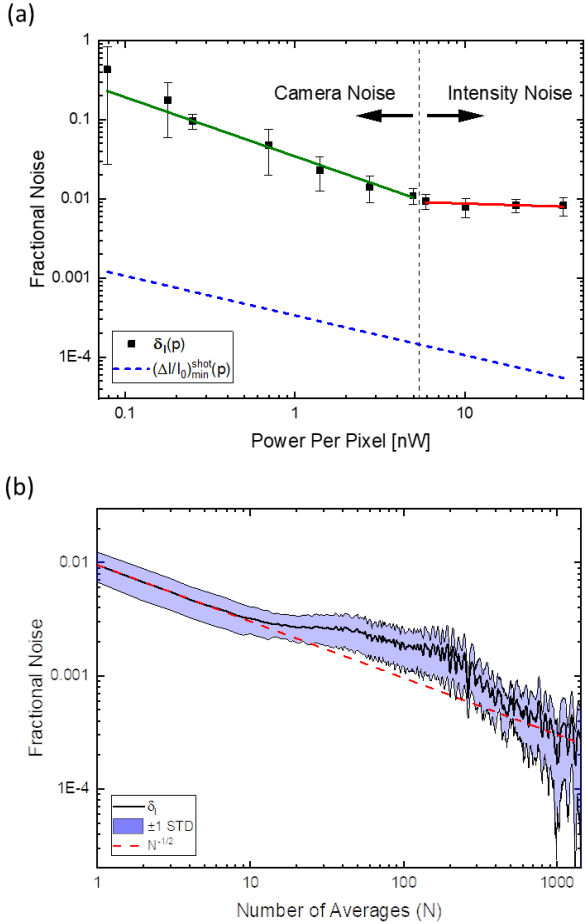

**Fig. 4.** Spectrometer sensitivity. (a) The fractional noise as a function of power per pixel. The noise figure is much higher than the shot noise limit shown in the blue dotted line. When the incident power per pixel is lower than 5 nW, the camera noise is dominant. While above that power, intensity noise is dominant. (b) Fractional noise as a function of average time with the total input power of about 3 nW. $\delta_I \approx 3 \times 10^{-4}$ per detection channel is achieved with 1000 averages.

## 3. CONCLUSION

In summary, we have developed a frequency comb spectrometer in the long-wave infrared region (8-10 μm) using an immersion grating. The highest frequency resolution of 463 MHz for dispersive spectrometer is achieved. In our current implementation, the frequency resolution of the spectrometer is limited by the size of the immersion grating. However, the resolution can be significantly increased by performing comb mode-resolved spectroscopy demonstrated in this work and stepping the repetition rate of the comb, which will push the limit of resolution to the linewidth of individual comb teeth.

The next application of this spectrometer is to perform cavity-enhanced time-resolved spectroscopy, which will exploit high sensitivity for the detection of transient molecules like $NO_3$, Criegee intermediates, and carbonic acid for applications in biology, geology, and atmospheric science, with time resolution of about 1 μs limited by the integration time of a camera. Moreover, we plan to incorporate the buffer gas cooling and molecular beam techniques with this spectrometer to perform quantum state-resolved spectroscopy of cold, large molecules like $C_{60}$ and dimethyl ether.

**Funding**.

K. I. is supported by the JSPS Overseas Research Fellowships. This work was supported by AFOSR Grant No. FA9550-15-1-0111, the DARPA SCOUT Program, NIST, and NSF PHYS-1734006.